\documentclass{icga}
\usepackage{graphicx}
\newcommand{\BibTeX}{{\rm B\kern-.05em{\sc i\kern-.025em b}\kern-.08em
    T\kern-.1667em\lower.7ex\hbox{E}\kern-.125emX}}

\title{AVOIDING ROTATED BITBOARDS WITH DIRECT LOOKUP}
\runningtitle{ICGA}
\author{Sam Tannous\thanks{email:sam.tannous@gmail.com}}
\affiliation{Durham, North Carolina, USA}
\issue{March, 2007}
\usepackage{fancyvrb}
\usepackage{listings}
\lstdefinestyle{numbers} 
{numbers=left, 
  stepnumber=1, 
  numberstyle=\scriptsize\sffamily,numbersep=10pt} 

\begin{document}
\maketitle
\begin{abstract}
This paper describes an approach for obtaining direct access to the attacked
squares of sliding pieces without resorting to rotated
bitboards. The technique involves creating four hash tables using the
built in hash arrays from an interpreted, high level language.  
The rank, file, and diagonal occupancy are first isolated by masking
the desired portion of the board.  The attacked squares are then directly retrieved 
from the hash tables.
Maintaining incrementally updated rotated bitboards
becomes unnecessary as does all the updating, mapping and shifting
required to access the attacked squares.  Finally, rotated bitboard
move generation speed is compared with that of the direct hash table lookup 
method.
\end{abstract}

\section{Introduction}
Prior to the their introduction by the Soviet chess program
KAISSA in the late 1960s, bitboards were used in checkers
playing programs as described in \citeaby{samuel59}.  
The elegance and performance advantages of bitboard-based 
programs attracted many chess programmers and bitboards were used by 
most early programs (\citeaby{slate78}, \citeaby{bitman70},
and \citeaby{hyatt90}).
But to fully exploit the performance advantages of parallel, bitwise 
logical operations afforded by bitboards, most programs maintain,
and incrementally update, rotated bitboards.  These rotated bitboards allow
for easy attack detection without having to loop over the squares
of a particular rank, file, or diagonal as described in \citeaby{heinz97}
and \citeaby{hyatt99}.  The file occupancy is computed by using an
occupancy bitboard rotated 90 degrees and then using the rank attack
hash tables to find the attacked squares.  Once the attacked squares are known, they are
mapped back to their original file squares for move generation.
The diagonal attacks are handled similarly except that the rotation
involved is 45 (or -45) degrees depending on which diagonal is being
investigated.   These rotated occupancy bitboards are incrementally 
updated after each move to avoid the performance penalty of dynamically
recreating them from scratch at every move.

\section{Direct Lookup}
As researchers and practitioners explore Shannon type B 
approaches to chess programming \citebay{shannon50}, code clarity 
and expressive power become important in implementing complex
evaluation functions and move ordering algorithms. 
Many high level programming languages (notably Python \citebay{python93}) have 
useful predefined data structures (e.g. associative arrays) which are dynamically
resizable hash tables that resolve collisions by probing techniques.  The basic
lookup function used in Python is based on Algorithm D: Open Addressing with 
Double Hashing from Section 6.4 in \citeaby{knuth98}.  We define four
dictionaries that are two dimensional
hash tables which the are main focus of this paper: 
\textbf{\textsf{rank\_attacks}}, 
\textbf{\textsf{file\_attacks}}, 
\textbf{\textsf{diag\_attacks\_ne}},
and \textbf{\textsf{diag\_attacks\_nw}} 
representing the rank, file, and two diagonal directions (``ne'' represents
the northeast A1-H8 diagonals and ``nw'' represents the northwest A8-H1 
diagonals).  In order to use these hash tables directly, we need to also
create rank, file and diagonal mask bitboards for each of the squares
(e.g. $diag\_mask\_ne[c4] = a2 | b3 | c4 | d5 | e6 | f7 | g8$).
These hash tables only need to be
generated at startup.  The initial cost of calculating these tables 
can be avoided altogether if the table
values are stored in a file and simply retrieved.  


\begin{figure}[htbp] 
  \begin{center}
    \includegraphics[scale=0.5]{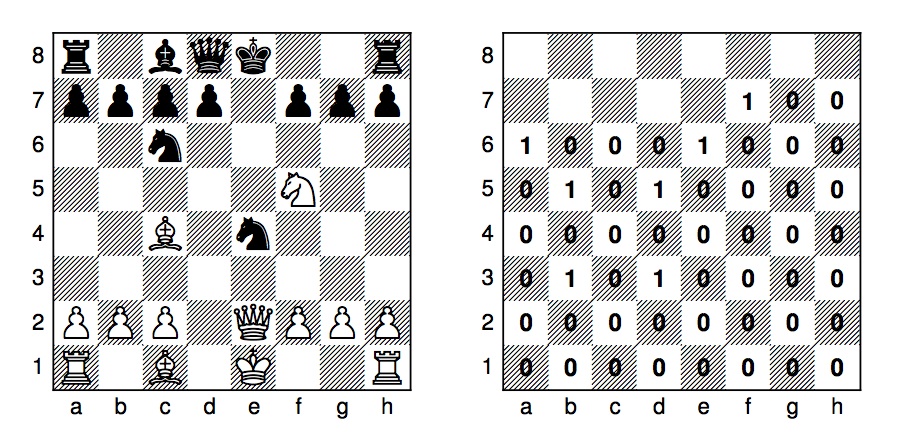}
\caption{C4 White Bishop Attacks and Attacked Squares Bitboard}
\label{diagattack}
\end{center}
\end{figure}

The first dimension represents the 
location of the attacking (sliding) piece and the second dimension represents
the occupancy of the particular rank, file, or diagonal.  The first dimension has
64 possible values and the second has 256 possible values (except for the
diagonals with fewer then eight squares).  While
the sizes of these hash tables are small, the actual values are fairly large
(up to $2^{64}-1$).  The reason for this is that these hash tables are called
directly from the bitboard values retrieved from the chess board. 
In Figure \ref{diagattack}, the squares attacked by the bishop at square c4
would ideally be found by simply calculating the occupancy of the two diagonals
intersecting at the square c4 and then performing a logical OR of the attacked
squares provided by direct lookup of the two diagonal hash tables
and then removing squares occupied by friendly pieces.
The techniques described in this paper provide the attacked squares 
that are both unoccupied and occupied.  These same attack
vectors are also used in evaluation functions that require attacks from
a certain square as well as attacks on a certain square.

\subsection{Rank Attacks}
The rank attack hash array can best be understood by starting with the
first rank.  (Note: in the subsequent listings, the convenience variables
for each square are created so that h1=1,
a1=128, h8=72057594037927936, a8=9223372036854775808, etc.)
The rank attack for the first rank is given by the following:

$$ rank\_attacks_{rank_1}(p_{rank_1},o_{rank_1}) = \sum _{i=l}^{p_{rank_1}-1} B_i + \sum _{i=p_{rank_1}+1}^{r}B_i $$

\noindent
where $p_{rank_1}$ is the position of the sliding piece (rook or queen) on the 
first rank,
$o_{rank_1}$ is the occupancy value for the first rank, $l$ is the first occupied
square to the left of the sliding piece, and $r$ is the first occupied
square to the right of the sliding piece.  And $B_i$ is the value given 
by 

$$ B_i = \cases{2^i,&if 1 exists at $i^{th}$ bit of $o_{rank_1}$; \cr
                0,&otherwise. \cr} $$

\noindent
Then finally, to find the rank attacks at the $i^{th}$ rank, we
simply move this first rank value ``up'' in rank by multiplying 
by $256^{rank-1}$ since moving a piece up one rank on the chessboard  
is the same as left shifting a binary number by 8 or multiplying by $2^8$.

$$ rank\_attacks_{rank_i}(p_{rank_i},o_{rank_i}) = rank\_attacks_{rank_1}(p_{rank_1},o_{rank_1}) \cdot 256^{i-1} $$

\noindent
where the piece position index and occupancy index at rank $i$ are also
multiplied by the same value as the attack

$$ p_{rank_i} = p_{rank_1} \cdot 256^{i-1}$$
$$ o_{rank_i} = o_{rank_1} \cdot 256^{i-1}$$

An implementation of this is shown in Listing \ref{rank}.  
Here, the function's outer loop (variable i in line 3) 
iterates the attacking piece over the squares of
the first rank beginning with square h1.  The rank\_attacks hash table
is initialized in line 2 and in lines 4 and 5.  The second loop iterates over
the possible 256 occupancy values for a rank (line 6).  After some initialization,
the function moves one square to the right of the attacking piece,
adding the value to the hash table.  If the square is occupied, there is a
piece that will block further movement in this direction and so we break out of 
this right side summation.  This process is repeated for the left side of the 
attacking square (lines 12-15).  Finally, when the rank\_attack hash table is complete
for the particular attacking square, the function shifts the values for each 
respective rank for the remaining ranks (lines 16-21).  Note that this
hash table includes blocking squares that are occupied by both enemy
and friendly pieces.  The friendly piece occupancy will need to be 
removed before assembling the legal moves.

\begin{lstlisting}[float=ht,frame=tb,label=rank,caption={Rank Attack Lookup Table}]{}
def get_rank_attacks ():
    rank_attacks = {}
    for i in range(8):
        for r in range(8):
            rank_attacks[1 << (i + (r * 8))] = {}
        for j in range(256):
            rank_attacks[1 << i][j] = 0
            for right in range(i-1, -1, -1):
                rank_attacks[1 << i][j] |= 1<<right  # save it
                if ((1 << right) & j != 0):  # non empty space
                    break 
            for left in range(i+1,8):
                rank_attacks[1 << i][j] |= 1 << left  # save it
                if ((1 << left) & j != 0):  # non empty space
                    break
            for rank in range(1,8):
                x = 1 << (i+(rank*8))
                y = j << (rank*8)
                value = rank_attacks[1 << i][j]
                newvalue = value << (rank*8)
                rank_attacks[x][y] = newvalue 

    return(rank_attacks)
\end{lstlisting}

\subsection{File Attacks}
The file attacks hash table uses the values obtained in the rank attack
table on the first rank and performs a 90 degree rotation.  
In the approach shown here, the 8th file
$file\_attacks_{file_8}$ hash table is obtained by converting the rank 1 
$rank\_attacks_{rank_1}$
table to base 256.  The bitboard position of the sliding piece 
as well as the occupancy are also converted in a similar fashion.

$$ file\_attacks_{file_8}(p_{file_8},o_{file_8}) = \sum_{i=1}^{8} B_i \cdot 256^i $$

where $B_i$ is the $i^{th}$ bit of the rank 1 rank\_attacks table 
(with h1 being the LSB and a1 being the MSB) and

$$ p_{file_8} = p_{rank_1} \cdot 256^{(9- \tilde f)} $$
$$ o_{file_8} = \sum_{i=1}^{8} O_{file_i} \cdot 256^i$$

where $\tilde f$ is the actual file number of the position square $p_{file_8}$
on the first rank and 
$O_{file_i}$ is the $i^{th}$ bit of the occupancy on the first
rank.

\begin{lstlisting}[float=tp,frame=tb,label=file,caption={File Attack Lookup Table}]{}
def get_file_attacks ():
    # this routing assumes that the rank_attacks have already
    # been calculated.
    file_attacks = {}
    for i in range(64):
        r = rank[1 << i] - 1
        mirror_i = rank_to_file((1 << i) >> (8*r)) << r
        file_attacks[mirror_i] = {}
        for j in range(256):
            mirror_j = rank_to_file(j) << r
            value = rank_attacks[1 << i][j << (8*r)]
            lower_value = value >> (8*r)
            file_value = rank_to_file(lower_value)
            final_value = file_value << r
            file_attacks[mirror_i][mirror_j] = final_value
    return(file_attacks)
\end{lstlisting}

The implementation of this is shown in Listing \ref{file} and uses the rank\_attacks
hash table found earlier (line 11).  This function has 
an outer loop that ranges over the 64 squares, for the attacking piece, 
and for each of these, an inner loop that loops over all the occupancy values.  
In line 7, we find the symmetric square
value if reflected across the A8-H1 diagonal (e.g. g1 is reflected across
the line of symmetry and onto square h2, f1 to h3, etc.).  In this way, the 
position values are flipped or ``rotated'' 90 degrees and the occupancy 
values are also rotated in line 10.  The function rank\_to\_file() performs
this rotation by converting the number to base two and then to base 256.
In line 11, the attacked squares
that were calculated in Listing \ref{rank} are also rotated. 

\subsection{Diagonal Attacks} \label{diags}
The attacked squares along the diagonals
are a little more complex to calculate using the base conversion
technique used on the file\_attacks.  A more direct approach
like the one used to find the rank\_attacks, involving shifting and adding,
is used.  The diagonal hash tables can be found by summing over the squares
up to and including the blocking square.  The A1-H8 diagonal can be found 

$$ diag\_attacks\_ne(p,o) = \sum _{i=l}^{p-1} B_i + \sum _{i=p+1}^{r}B_i $$

\noindent
where $p$ is the position of the sliding piece (bishop or queen),
$o$ is the occupancy value for the diagonal, $l$ is the first occupied
square along the diagonal to the left of the sliding piece, and $r$ is the first occupied
square along the diagonal to the right of the sliding piece.  $B_i$ is the 
value of the number if the $i^{th}$ bit is set

$$ B_i = \cases{2^i,&if 1 exists at $i^{th}$ bit of $o$; \cr
                0,&otherwise. \cr} $$

The other diagonal hash table (for the A8-H1 direction) 
is not shown but has a similar structure.

\begin{lstlisting}[float=tp,frame=tb,label=diagslisting,
  caption={Generalized Attack Lookup Table}]{}
def get_attacks (square_list=None):
    attack_table = {}
    attack_table[0] = {}
    attack_table[0][0] = 0
    for i in range(len(square_list)):
        list_size = len(square_list[i])
        for current_position in range(list_size):
            current_bb = square_list[i][current_position]
            attack_table[current_bb] = {}
            for occupation in range(1 << list_size):
                moves = 0
                for newsquare in range(current_position+1,list_size):
                    moves |= square_list[i][newsquare]
                    if ((1 << newsquare) & occupation):
                        break
                for newsquare in range(current_position-1,-1,-1):
                    moves |= square_list[i][newsquare]
                    if ((1 << newsquare) & occupation):
                        break
                temp_bb = 0
                while (occupation):
                    lowest = lsb(occupation)
                    temp_bb |= square_list[i][bin2index[lowest]] 
                    occupation = clear_lsb(occupation)
    return(attack_table)

def get_diag_attacks_ne ():
    diag_values = [[h1],
                 [h2,g1],
                [h3,g2,f1],
               [h4,g3,f2,e1],
              [h5,g4,f3,e2,d1],
             [h6,g5,f4,e3,d2,c1],
            [h7,g6,f5,e4,d3,c2,b1],
           [h8,g7,f6,e5,d4,c3,b2,a1],
            [g8,f7,e6,d5,c4,b3,a2],
             [f8,e7,d6,c5,b4,a3],
              [e8,d7,c6,b5,a4],
               [d8,c7,b6,a5],
                [c8,b7,a6],
                 [b8,a7],
                  [a8]]
    return(get_diag_attacks(diag_values))

def get_diags_attacks_nw ():
    diag_values = [[a1],
                 [b1,a2],
                [c1,b2,a3],
               [d1,c2,b3,a4],
              [e1,d2,c3,b4,a5],
             [f1,e2,d3,c4,b5,a6],
            [g1,f2,e3,d4,c5,b6,a7],
           [h1,g2,f3,e4,d5,c6,b7,a8],
            [h2,g3,f4,e5,d6,c7,b8],
             [h3,g4,f5,e6,d7,c8],
              [h4,g5,f6,e7,d8],
               [h5,g6,f7,e8],
                [h6,g7,f8],
                 [h7,g8],
                  [h8]]
    return(get_diag_attacks(diag_values))
\end{lstlisting}

\begin{lstlisting}[float=tp,frame=tb,label=newrankfile,
  caption={Generalized Rank and File Attack Lookup Table}]{}
def get_rank_attacks ():
    # these are the rank square values
    rank_values = [[a1,b1,c1,d1,e1,f1,g1,h1],
                   [a2,b2,c2,d2,e2,f2,g2,h2],
                   [a3,b3,c3,d3,e3,f3,g3,h3],
                   [a4,b4,c4,d4,e4,f4,g4,h4],
                   [a5,b5,c5,d5,e5,f5,g5,h5],
                   [a6,b6,c6,d6,e6,f6,g6,h6],
                   [a7,b7,c7,d7,e7,f7,g7,h7],
                   [a8,b8,c8,d8,e8,f8,g8,h8]]
    return(get_attacks(rank_values))

def get_file_attacks ():
    # these are the file square values
    file_values = [[a1,a2,a3,a4,a5,a6,a7,a8],
                   [b1,b2,b3,b4,b5,b6,b7,b8],
                   [c1,c2,c3,c4,c5,c6,c7,c8],
                   [d1,d2,d3,d4,d5,d6,d7,d8],
                   [e1,e2,e3,e4,e5,e6,e7,e8],
                   [f1,f2,f3,f4,f5,f6,f7,f8],
                   [g1,g2,g3,g4,g5,g6,g7,g8],
                   [h1,h2,h3,h4,h5,h6,h7,h8]]
    return(get_attacks(file_values))
\end{lstlisting}

An implementation of this is shown in Listing \ref{diagslisting}.    Each diagonal is looped
over (line 5) for the outer loop and the attacking piece is moved along the diagonal
for the inner loop (line 7).  For each position of the attacking piece, all of the
possible occupancies are generated (line 10) and the two inner loops, one for the
right side (lines 12-15) and one for the left side (lines 16-19), are used to accumulate
open squares until blocking bits are encountered.  The function completes by 
converting the occupancy value to a bitboard number along the diagonal.
Not shown is a hash table called \textbf{bin2index} 
used to convert bitboard values to square index
values (e.g. a1$\rightarrow$7).
The function is called with a list of lists of the values of the diagonals.
For the A1-H8 direction (also referred to as the ``northeast'' or ne direction),
the diagonal values are shown in lines 28-42 and for the A8-H1 diagonals, the
diagonal values passed into the function are shown in lines 46-60.

This algorithm is general enough to allow for the rank and file attack
tables to be generated and these are reformulated to work with this 
approach and the listings are shown in Listing \ref{newrankfile}.

\section{Experimental Results} \label{resultssection}
\begin{table}[h]
\begin{center}
\begin{tabular}{|l|c|c|}
\hline OS and CPU & Rotated Bitboards Time (s) & Direct Lookup Time (s)  \\
\hline 
\hline 
OS X 10.4.9 2.33 GHz Intel Core 2 Duo & 7.29   & 6.42  \\
Linux 2.6.9 3.4 GHz Intel Quad Xeon   & 8.82  & 7.67 \\
OS X 10.4.9 1.67 GHz PowerPC G4       & 16.31  & 14.13  \\
SunOS 5.8 1.5 GHz dual UltraSPARC-IIIi     & 23.95   & 19.06  \\
FreeBSD 6.2 500 MHz Pentium 3         & 58.17  & 44.85 \\
\hline
\end{tabular}
\caption{Move Generation Results for Rotated Bitboards and Direct Lookup. \label{results}}
\end{center}
\end{table}

A performance comparison of simple move generation 
was made between a rotated bitboard implementation and
a direct lookup implementation.  The test results are shown in Table \ref{results}.
The times shown reflect a comparison of the move generation routines.
A well known and well studied set of test cases exists in the Encyclopedia of Chess
Middle Games (ECM).
Positions were selected from the ECM \citebay{krogius80} and for each of the 879
test positions, a list of the main board position as well as three rotated boards
were precalculated and saved in a list used by both methods.  The moves were then generated 
for each of these 879 positions using the same list of bitboards generated earlier.
The move generation functions for direct lookup
and those for the rotated bitboards differed only in how they handled the sliding piece attacks.
This process was repeated 10 times for each of the two types of approaches.
In generating moves for rotated bitboards, we required additional shifting and masking operations
before the lookup of the attacks could take place.  Furthermore, the overhead
of maintaining and updating the rotated bitboards is not accounted for since
these test positions represent games in mid play where the rotated bitboards
were precalculated. 

The results shown indicate that directly looking up the attacking moves
for sliding pieces in hash tables improves the move generation speeds 
from 10\% to 15\%  depending on computer architecture.  Further efficiencies
can be expected in a full implementation where the overhead of maintaining
rotated bitboards is eliminated.
The implementation and test code is made available in an Open-Source,
interactive, chess programming module called ``Shatranj'' \citebay{shatranj06}. 

\section{Conclusions and Future Work}
We have described an approach for obtaining direct access to the attacked
squares of sliding pieces without resorting to rotated
bitboards. Detailed algorithms and working code illustrate how  
the four hash tables were derived.  The attacked squares are directly retrieved 
from the hash tables once the occupancy for the particular rank, file or diagonal
was retrieved by the appropriate masks.  Using these four hash tables, 
maintaining incrementally updated rotated bitboards
becomes unnecessary as does the required shifting and masking 
required to obtain the consecutive occupancy bits for a rank, 
file or diagonal.
In addition to simplifying the implementation, we can expect a performance
improvement in move generation of at least 10\%.  
Taking the implementation a bit further,  
the hash tables described in this
paper are also useful for implementing evaluation functions
which include piece mobilty and attack threats.  When the additional
impact of complex evaluation functions is taken into account, 
the speed improvements should be greater then the results presented here.

Since most chess implementations do not use a high level interpreted language
such as Python,  it is difficult to estimate the effect of cache loading and
execution speed.  The results presented here only reflect the savings
seen by move generation and not those of a fully implemented chess engine.
Further research is needed to quantify the effect of these changes on 
cache utilization in a complete chess engine.

Alternatives to rotated bitboards have gained some popularity recently.
Minimal perfect hash functions as descibed in 
\citebay{czech92} have been used to create hash tables where the
index is calculated based on the mover square and occupancy bitboard.
A recent refinement of this described in \citebay{leiserson98},
called magic move generation, further reduces the memory requirements of the
hash table.   In this approach, a ``magic multiplier'' for a particular 
square is multiplied by an occupancy bitboard and then shifted by
another ``magic'' number.  This provides a hash index where the attacked
squares can be retrieved from a hash table.  Performance comparisons
of the built in hash tables provided by interpreted languages and
techniques involving manually creating minimal perfect hash functions
as well as hash functions using de Bruijn sequences (a.k.a. magic move
generation techniques) could also be explored in future work.  
Representation of chess knowledge with the data structures
provided by high level languages seems to have received very little
attention since the primary focus of the majority of work has been
improving execution speed, an area that places interpreted languages at a 
distinct disadvantage.

\bibliography{avoiding-rotations}

\end{document}